\documentclass[%
 aip,
 amsmath,amssymb,
 preprint,%
]{revtex4-1}

\usepackage{graphicx}
\usepackage{dcolumn}
\usepackage{bm}
\usepackage[utf8]{inputenc}
\usepackage[T1]{fontenc}
\usepackage{mathptmx}
\usepackage{etoolbox}
\usepackage{float}
\usepackage{hyperref}
\usepackage{amsmath,amssymb,amsfonts}
\usepackage{algorithmic}
\usepackage{graphicx}
\usepackage{textcomp}
\usepackage{xcolor}
\usepackage{bm}
\usepackage{tabularx}
\usepackage{booktabs}
\usepackage{array}
\usepackage[table]{xcolor}

\usepackage{booktabs}
\usepackage{placeins} 
 
\usepackage{booktabs}
\usepackage[table]{xcolor}
\definecolor{apricot}{RGB}{251,206,177}
\usepackage{array}
\newcolumntype{C}[1]{>{\centering\arraybackslash}m{#1}}

\makeatletter
\def\@email#1#2{%
 \endgroup
 \patchcmd{\titleblock@produce}
  {\frontmatter@RRAPformat}
  {\frontmatter@RRAPformat{\produce@RRAP{*#1\href{mailto:#2}{#2}}}\frontmatter@RRAPformat}
  {}{}
}%
\makeatother
\begin{document}

\preprint{}

\title[]{ALD W-Doped SnO$_2$ TFTs for Indium-Free BEOL Electronics}
\author{Mansi Anil Patil}
\author{Devarshi Dhoble}%
\author{Shivaram Kubakaddi} 
\author{Mamta Raturi}
\affiliation{ 
Department of Electrical Engineering, Indian Institute of Technology Bombay, Mumbai 400076, India 
}%
\author{Marco A Villena}
\affiliation{ 
Department of Electronics and Computer Technology, Faculty of Sciences, University of Granada, Fuentenueva Avenue s/n, 18071 Granada, Spain}%
\author{Gaurav Thareja}
\affiliation{ 
Applied Materials Inc, Santa Clara, California, USA
}%
\author{Saurabh Lodha} 
 \email{slodha@ee.iitb.ac.in}
\affiliation{ 
Department of Electrical Engineering, Indian Institute of Technology Bombay, Mumbai 400076, India 
}%


\date{\today}

\begin{abstract}
This work reports back-end-of-line (BEOL) compatible, thin-film transistors (TFTs) with sub-10 nm tungsten-doped tin oxide (TWO) channels deposited by atomic layer deposition (ALD) at 150 $^\circ$C. TFTs with undoped SnO$_{\mathrm{x}}$, undoped WO$_{\mathrm{x}}$, and W-doped SnO$_{\mathrm{x}}$ channels with W concentrations of 5\% and 10\% were investigated. TFT with 10\% W doping exhibited the best electrostatic control and overall device performance. Post-fabrication O$_{\mathrm{2}}$ annealing at 300 $^\circ$C for 5 minutes significantly enhanced device characteristics, reducing the subthreshold swing (SS) by nearly 2x, increasing the I$_{\mathrm{on}}$/I$_{\mathrm{off}}$ ratio from 10$^7$ to 10$^9$, decreasing hysteresis by nearly 3x and positive bias stress-induced threshold shift by over 2x to a low value of 93 mV at a stress field of 4 MV/cm.  Kinetic Monte Carlo simulations using Ginestra$^{\mathrm{TM}}$ support the experimental observations and attribute the bias instability to charge trapping in the gate dielectric and at the interface. This work demonstrates low-temperature ALD-grown TWO TFTs as a promising indium-free platform for BEOL and monolithic 3D integration.
\end{abstract}

\maketitle
%

\section*{Introduction}
\vspace{-15pt}
Due to the unique combination of high electron mobility, optical transparency, low off-current, large-area uniformity, and low-temperature as well as low-cost fabrication, thin-film transistors (TFTs) employing amorphous oxide semiconductor (AOS) channels have gained significant interest over the last two decades.\cite{ref1,ref2,ref3,ref4} They are especially desirable for applications in the flexible electronics and next-generation flat panel displays, along with low thermal budget back-end-of-line (BEOL) and monolithic 3D integration.\cite{ref5,ref6,ref7} AOS materials incorporate heavy post-transition metal cations with (n$-$1)d$^{\mathrm{10}}$ns$^{\mathrm{0}}$ (n $\geq$ 4) electronic configurations, which allow for a conduction band with isotropic, spherical, and spatially extended s-orbitals in contrast to the anisotropic sp$^{\mathrm{3}}$ orbitals in Si or other conventional semiconductors. This leads to relatively high electron mobility even in amorphous channels.\cite{ref4,ref1} In$_{\mathrm{2}}$O$_{\mathrm{3}}$ and In$_{\mathrm{2}}$O$_{\mathrm{3}}$-based materials (indium-gallium-zinc oxide (IGZO), indium-zinc oxide (IZO), indium-tin oxide (ITO), indium-tungsten oxide (IWO), etc.) are the most widely researched and benchmarked amorphous oxide semiconductors for high-performance TFT applications.\cite{ref8,ref9,ref37} Indium is a critically scarce resource, occurring at an average abundance of only 0.05 ppm in the earth’s crust,\cite{ref10} and its consumption has risen sharply in recent years (around 120\% increase from 2020 to 2024).\cite{ref11} This necessitates the development of viable alternatives to indium-based amorphous oxide semiconductors.

Tin dioxide (SnO$_{\mathrm{2}}$) can be a promising indium-free channel semiconductor due to its advantageous electronic structure, with Sn$^{4+}$(4d$^{10}$5s$^0$) cations providing s-orbital-dominated conduction similar to In$^{3+}$, paired with its earth abundance and cheaper cost.\cite{ref12} Undoped SnO$_{\mathrm{2}}$ has a high intrinsic carrier density, leading to negative threshold voltages (V$_{\mathrm{th}}$) and high off-state leakage currents (I$_\mathrm{off}$), resulting in high power consumption.\cite{ref13,ref14} Both physical vapor deposition (PVD) and atomic layer deposition (ALD) techniques have been explored for the deposition of undoped SnO$_{\mathrm{2}}$.\cite{ref15,ref16} However, these reports mainly demonstrate polycrystalline films, and obtaining non-polycrystalline (or fully amorphous) layers remains difficult. This limitation can lead to device-to-device performance variability. Several studies have explored introducing different dopants into SnO$_{\mathrm{2}}$ to address these challenges. For example, Zr,\cite{ref17} Hf,\cite{ref14} Zn,\cite{ref18} Si,\cite{ref19,ref20} Ni,\cite{ref21} Ga,\cite{ref22} Ti,\cite{ref23} and Mg\cite{ref21} have been investigated. These studies show that both aliovalent and isovalent cations can help reduce excess carriers and modify trap states. However, trade-offs still remain; for example, solution-based approaches are cost-effective but they typically require high-temperature post-annealing ($\geq$ 500 $^\circ$C), which limits compatibility with BEOL processes.\cite{ref13,ref23,ref24} They also suffer from poor film uniformity at larger length scales and high defect densities. PVD techniques like sputtering can lower the thermal budget. But sputtered films often face challenges in maintaining consistent stoichiometry, are typically used for  relatively thick channels, and struggle with uniformity over high-aspect-ratio (HAR) structures. In one study, W-doped SnO$_{\mathrm{2}}$ TFTs were fabricated by sputtering a SnO$_{\mathrm{2}}$ target coated with tungsten flakes.\cite{ref25} It demonstrated that W doping effectively suppresses carrier concentration as well as crystallinity to get controlled amorphous films. However, this approach offers limited control over the W dopant concentration and spatial uniformity.

In this study, we report sub-10 nm tungsten-doped tin oxide (TWO) films deposited by ALD,  ensuring conformal, highly uniform, and reproducible deposition. TWO TFTs with varying W concentration were fabricated with 6.8 nm thick channels, a 10 nm ALD HfO$_{\mathrm{2}}$ gate dielectric, and a channel length (L$_{\mathrm{ch}}$) of 5 $\mu$m. The device geometry was scaled significantly when compared to similar reports (Table 1).\cite{ref15,ref16,ref20,ref25} Best transistor gate control was achieved for 10\% W doping of the TWO channel. O$_{\mathrm{2}}$ rapid thermal annealing at 300 $^\circ$C for 5 mins significantly improved the device parameters and overall performance. Additionally, positive bias stress (PBS) measurements on annealed TWO devices showed better PBS stability compared to unannealed TWO devices, as well as  previously reported doped SnO$_{\mathrm{x}}$ TFTs.
\section*{Experimental}
\vspace{-15pt}
\begin{figure*}[htbp]
\centering
\includegraphics[width=0.8\textwidth]{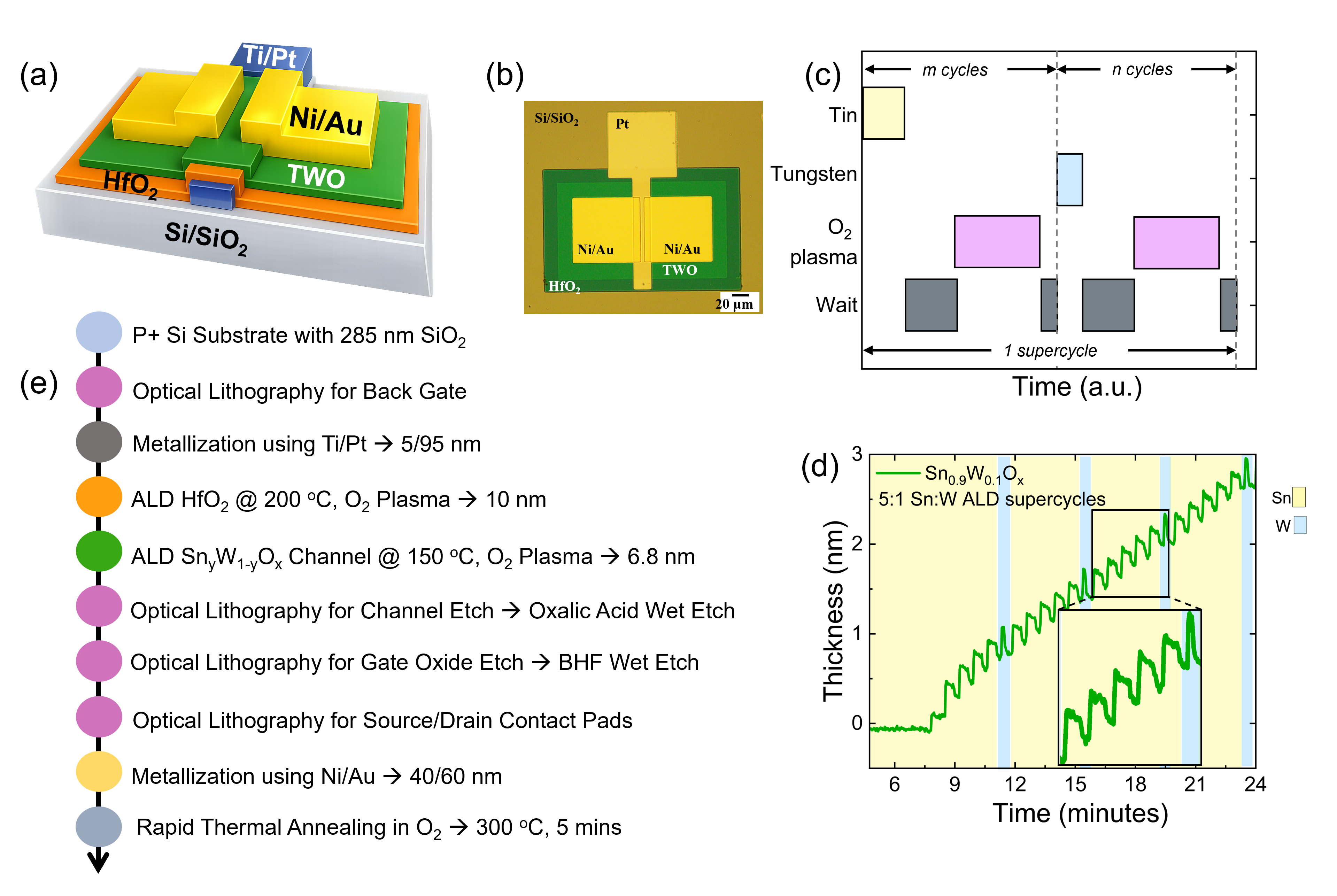}
\caption{\label{fig:Fig. 1}(a) 3D schematic illustration of the device architecture. (b) Optical microscope image
of the fabricated TFT. (c) Schematic of an ALD supercycle sequence for fabricating tungsten-doped tin oxide, Sn$_{\mathrm{x}}$W$_{\mathrm{1-x}}$O$_{\mathrm{y}}$, where an m:n (SnO$_{\mathrm{x}}$:WO$_{\mathrm{x}}$) ALD cycle ratio constitutes one supercycle. (d) In-situ spectroscopic ellipsometry growth profile of Sn$_{\mathrm{0.9}}$W$_{\mathrm{0.1}}$O$_{\mathrm{x}}$ thin film, showing periodic Sn (yellow regions) and W (blue regions) subcycles, indicating a 5:1 Sn:W ratio. (e) Process flow for the TFT fabrication.}
\end{figure*}
The TFTs were fabricated on p+ Si/SiO$_{\mathrm{2}}$ substrates (285 nm thick oxide), which were cleaned sequentially with acetone and IPA. A 3D device schematic and top-view optical image of a final fabricated TFT are shown in Fig.~\ref{fig:Fig. 1}(a) and Fig.~\ref{fig:Fig. 1}(b), respectively. Back-gate (BG) electrodes were defined by an optical lithography system, microwriter from Durham Magneto Optics Ltd., followed by sputtering of a Ti/Pt bilayer (5/95 nm) in an AJA sputter system and lift-off. A 10 nm thick HfO$_{\mathrm{2}}$ gate dielectric was then deposited by plasma-enhanced ALD (PEALD) in a Veeco FIJI G2 system at 200 $^\circ$C using tetrakis(dimethylamino)hafnium(IV) (TDMAH) precursor and oxygen plasma. Four variants of channel material were deposited to study the effect of tungsten doping: undoped SnO$_{\mathrm{x}}$, undoped WO$_{\mathrm{x}}$, 5\% W-doped SnO$_{\mathrm{x}}$ (Sn$_{\mathrm{0.95}}$W$_{\mathrm{0.05}}$O$_{\mathrm{x}}$, using a 10:1 Sn:W ALD cycle), and 10\% W-doped SnO$_x$ (Sn$_{\mathrm{0.9}}$W$_{\mathrm{0.1}}$O$_{\mathrm{x}}$, using a 5:1 cycle). The schematic of an ALD supercycle of W and Sn is depicted in Fig.~\ref{fig:Fig. 1}(c). Step-wise thickness evolution of the channel was monitored using in-situ spectroscopic ellipsometry in the ALD chamber, where each time a W subcycle was introduced, the anticipated thickness staircase with small plateaus was observed, confirming successful W incorporation (Fig.~\ref{fig:Fig. 1}(d)). Growth-per-cycle (GPC) values of 1.16 Å/cycle and 0.66 Å/cycle were measured for the individual SnO$_{\mathrm{x}}$ and WO$_{\mathrm{x}}$ films respectively. All channel films were deposited using O$_{\mathrm{2}}$ plasma at 150 $^\circ$C using tetrakis(dimethylamino)tin(IV) (TDMASn) precursor for Sn and bis(tert-butylimino)bis(dimethylamino)tungsten(VI) (BTBMW) precursor for W. Channel thickness of 6.8 nm was confirmed by in-situ ellipsometry (J.A. Woollam). Channels were patterned using optical lithography and etched in 0.5 M oxalic acid. Similarly, for dielectric isolation after lithography, HfO$_{\mathrm{2}}$ was etched using buffered HF (BHF). Source/drain (S/D) regions were defined using optical lithography followed by sputtering Ni/Au (40/60 nm) and lift-off. The fabricated devices were annealed in a rapid thermal process (RTP) tool at 300 $^\circ$C in an oxygen ambient for 5 mins. The overall process sequence is summarized in Fig.~\ref{fig:Fig. 1}(e). The devices were stored in a glovebox after fabrication and electrically characterized using an Agilent B1500A semiconductor parameter analyzer. X-ray photoelectron spectroscopy (XPS) measurements were performed in a PHI 5000 VersaProbe II system.
\section*{Results and Discussion}
\vspace{-15pt}
\begin{figure*}[htbp]
\centering
\includegraphics[width=\textwidth]{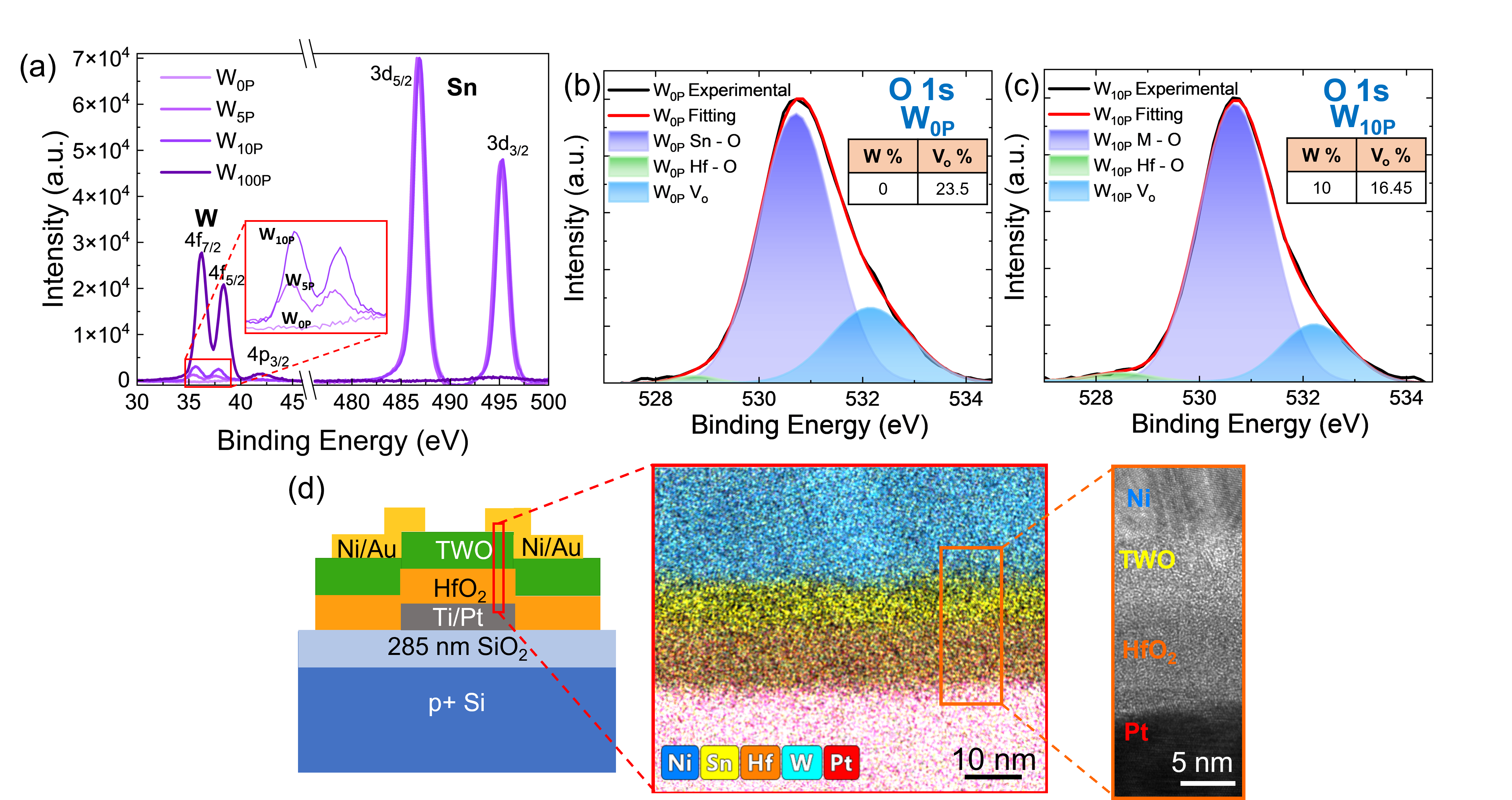}
\caption{\label{fig:Fig. 2}(a) XPS survey spectra of ALD W-doped SnO$_{\mathrm{x}}$ films, the inset shows a magnified view of the W 4f$_{\mathrm{7/2}}$ and W 4f$_{\mathrm{5/2}}$ peaks. (b) Deconvoluted O 1s XPS spectrum of the W$_{\mathrm{0P}}$ (SnO$_{\mathrm{x}}$) film; inset shows extracted oxygen vacancy concentration. (c) Deconvoluted O 1s XPS spectrum of the W$_{\mathrm{10P}}$ TWO film; inset shows extracted oxygen vacancy concentration. (d) Schematic and cross-sectional STEM-EDS mapping of the Ni/Au–TWO–HfO$_{\mathrm{2}}$–Pt/Ti device stack, showing distinct Ni, Sn, Hf, W, and Pt elemental distributions and sharp layer interfaces. }
\end{figure*}
For the four variations of channel material: W$_{\mathrm{0P}}$, W$_{\mathrm{5P}}$, W$_{\mathrm{10P}}$, and W$_{\mathrm{100P}}$, i.e., undoped SnO$_{\mathrm{x}}$, 5\% W-doped SnO$_{\mathrm{x}}$, 10\% W-doped SnO$_{\mathrm{x}}$, and undoped WO$_{\mathrm{x}}$, respectively, XPS characterization was performed to get a comparative analysis. Supplementary table S1 shows targeted and achieved W\% calculated using XPS analysis along with the supercycle ratio. Fig.~\ref{fig:Fig. 2}(a) shows normalized Sn 3d$_{\mathrm{5/2}}$ peaks for W$_{\mathrm{0P}}$, W$_{\mathrm{5P}}$, and W$_{\mathrm{10P}}$ in the survey scans to show comparative W\%. Fig.~\ref{fig:Fig. 2}(b) and Fig.~\ref{fig:Fig. 2}(c) show the deconvoluted, narrow scan O 1s data for W$_{\mathrm{0P}}$ and W$_{\mathrm{10P}}$ films deposited on HfO$_{\mathrm{2}}$, respectively. The O 1s spectra consist mainly of three peaks: M-O (oxygen combined with Sn and/or W) peak around 530.6 eV, V$_{\mathrm{o}}$ (oxygen vacancies along with surface contaminants and chemisorbed oxygen-related species like OH$^{-}$, CO, and CO$_{\mathrm{2}}$) at 532.1 eV,\cite{ref26} and a very small percentage from the HfO$_{\mathrm{2}}$ film at the bottom at 528.8 eV. V$_{\mathrm{o}}$ concentration decreased from 23.5\% to 16.45\% after doping the W$_{\mathrm{0P}}$ film with 10\% W. This is because of the strong interaction between tungsten and oxygen in the channel. Tungsten forms significantly stronger W–O (670 kJ/mol) bonds than Sn–O (531 kJ/mol), making oxygen atoms more tightly bound and less likely to form vacancies during deposition and annealing. In addition, the oxygen vacancy concentration can also drop because of the higher valence state of W$^{\mathrm{6+}}$ compared to Sn$^{\mathrm{4+}}$.\cite{ref27} The presence of W atoms thus improves the material's stability and effectively suppresses vacancy formation. Oxygen vacancies act as donor defects, and excess V$_{\mathrm{o}}$ contribute to high carrier concentration, resulting in high off-current of the transistor. Hence, controlling V$_{\mathrm{o}}$ is critical for controlling the electrical performance of TFTs.

The amorphous nature and structural characteristics of the TFT films were examined using cross-sectional transmission electron microscopy (TEM) on a fabricated device. The stack consists of Ni/TWO/HfO$_{\mathrm{2}}$/Pt from top to bottom as shown in Fig.~\ref{fig:Fig. 2}(d). The spatial distribution of Ni, Sn, W, Hf, and Pt throughout the stack is confirmed by elemental mapping using energy-dispersive x-ray spectroscopy (EDS). The TWO layer is confirmed to be around 6.8 nm thick, whereas the HfO$_{\mathrm{2}}$ layer has a consistent thickness of about 10 nm. High-resolution TEM images reveal that both HfO$_{\mathrm{2}}$ and TWO are amorphous and form smooth and clean interfaces. The absence of sharp Bragg reflections in the x-ray diffraction (XRD) spectrum (supplementary material Fig. S1) also confirms that the deposited TWO film is amorphous. The Pt/HfO$_{\mathrm{2}}$ (gate metal/dielectric) and Ni/TWO (S/D contact metal/channel) interfaces are smooth with low interfacial roughness and interlayer diffusion. Atomic force microscopy (AFM) analysis reveals a smooth surface (rms roughness < 0.15 nm) and a thickness of ~6.8 nm, consistent with the in-situ ellipsometry results (supplementary material Fig. S2).

\begin{figure*}[htbp]
\includegraphics[width=0.8\textwidth]{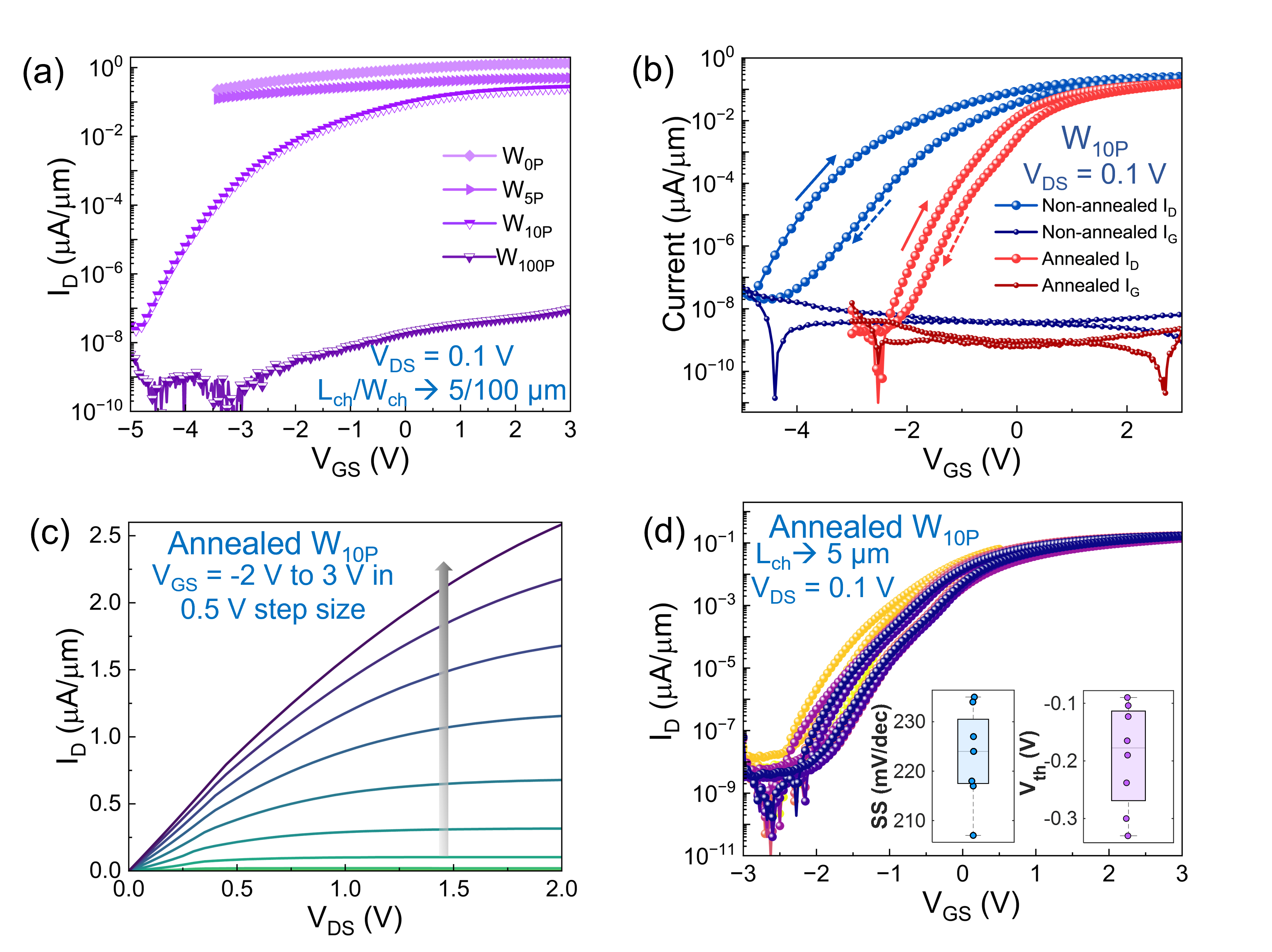}
\caption{\label{fig:Fig. 3}(a) Transfer characteristics (I$_{\mathrm{DS}}$ vs. V$_{\mathrm{GS}}$) of W-doped SnO$_{\mathrm{x}}$ (TWO) TFTs (at V$_{\mathrm{DS}}$ = 0.1 V) with varying tungsten doping concentration. (b) Transfer characteristics of the W$_{\mathrm{10P}}$ TWO device before and after annealing. (c) Output characteristics (I$_{\mathrm{DS}}$ vs. V$_{\mathrm{DS}}$) of the annealed W$_{\mathrm{10P}}$ TWO device for gate voltage varying from $-2$ V to +3 V in 0.5 V steps. (d) Device-to-device variability of transfer characteristics for the annealed W$_{\mathrm{10P}}$ TWO devices (L$_{\mathrm{ch}}$ = 5 $\mu$m), with inset box plots showing the distribution of subthreshold swing (SS) and threshold voltage (V$_{\mathrm{th}}$).}
\end{figure*}
Transfer characteristics (drain current I$_{\mathrm{DS}}$ vs. gate-to-source voltage V$_{\mathrm{GS}}$) of the four different channel compositions measured at fixed drain-to-source bias (V$_{\mathrm{DS}}$) of 0.1 V and L$_{\mathrm{ch}}$ of 5 $\mu$m are shown in Fig.~\ref{fig:Fig. 3}(a). The W$_{\mathrm{0P}}$ and W$_{\mathrm{5P}}$ devices do not exhibit significant modulation of the drain current by the gate voltage before the onset of gate leakage (I$_{\mathrm{GS}}$–V$_{\mathrm{GS}}$ curves in supplementary material Fig. S3). This is because the electron concentration in the channel is too high for gate modulation, and the off-state current for W$_{\mathrm{0P}}$ and W$_{\mathrm{5P}}$ in the negative gate voltage region is substantial, indicating high conductivity with little change in drain current.\cite{ref28,ref29} Even though the drain current drops in W$_{\mathrm{5P}}$ as compared to W$_{\mathrm{0P}}$, it doesn’t exhibit gate modulation, suggesting that 5\% W doping is not sufficient to suppress the excess carrier concentration. The W$_{\mathrm{100P}}$ device, on the other hand shows very low drain current (10$^{\mathrm{-8}}$ $\mu$A/$\mu$m or less) and poor gate modulation. This behavior results from the wide-bandgap n-type stoichiometric tungsten trioxide (WO$_{\mathrm{3}}$ and related WO$_{\mathrm{x}}$ phases) having a low density of mobile carriers and very low intrinsic conductivity.  WO$_{\mathrm{3}}$ tends to behave more like an insulating or resistive layer with low mobility and trap-limited transport, limiting field-effect performance.\cite{ref30,ref31,ref32} Importantly, the intermediate W$_{\mathrm{10P}}$ device exhibits significant gate modulation of drain current, displaying typical field-effect transistor behavior. The low I$_\mathrm{off}$ (2x10$^{-8}$ $\mu$A/$\mu$m) and an on/off current ratio (I$_\mathrm{on}$/I$_\mathrm{off}$) of around 10$^{7}$ show that 10\% W incorporation efficiently suppresses excess mobile carrier concentration while retaining sufficient mobile charge for conductivity modulation. Even though W$_{\mathrm{10P}}$ shows better gate modulation compared to the other three variants of film composition, it has poor electrical parameters because of interface and channel trap states. Open-air/furnace annealing at 300 $^\circ$C and higher for about 1 hour has been shown to enhance overall performance of SnO$_2$ TFTs by reducing defect densities.~\cite{ref25,ref19,ref34} RTP at 300 $^\circ$C in an oxygen ambient for 5 minutes significantly improved the subthreshold swing (SS, 2x, from 410 to 220 mV/dec), interface state density (D$_{\mathrm{it}}$, from $6.1 \times 10^{13}$ to $2.8 \times 10^{13}$ cm$^{-2}$eV$^{-1}$), I$_{\mathrm{on}}$/I$_{\mathrm{off}}$ (from 10$^{7}$ to 10$^{9}$), V$_{\mathrm{th}}$ (from $-1.09$ to $-0.02$ V), field-effect mobility ($\mu_{\mathrm{FE}}$, from 5.8 to 6.6 cm$^2$/V$\cdot$s) and hysteresis (from 0.83 to 0.31 V)  of the W$_{\mathrm{10P}}$ TFT (Fig.~\ref{fig:Fig. 3}(b)). The increase in $\mu_{\mathrm{FE}}$ is an under-estimate since it has not been corrected for the increase in contact resistance (R$_{\mathrm{c}}$) from 1.4 k$\Omega$ to 4.7 k$\Omega$ due to a reduction in the channel doping (less vacancies) caused by the O$_{\mathrm{2}}$ anneal (see supplementary material section S4 and Fig. S4). The O$_{\mathrm{2}}$ anneal passivates gate dielectric/channel interface states leading to an improved SS and $\mu_{\mathrm{FE}}$ due to reduced interface scattering. This short-time annealing process also reduces the overall thermal budget and is beneficial for system-level integration. Output characteristics (I$_{\mathrm{DS}}$-V$_{\mathrm{DS}}$) of the annealed W$_{\mathrm{10P}}$ transistor are plotted in Fig.~\ref{fig:Fig. 3}(c) for gate voltages varying from $-2$ to $+3$ V with a step size of $0.5$ V. They show well-defined, linearly increasing drain current at low V$_{\mathrm{DS}}$, followed by drain current saturation at higher bias, indicating ohmic S/D contacts and effective channel pinch-off. Fig.~\ref{fig:Fig. 3}(d) highlights the device-to-device variability in transfer characteristics for W$_{\mathrm{10P}}$ TFTs, showing a moderate spread in drain current, V$_{\mathrm{th}}$, and SS in multiple devices. The slight variability observed can be attributed to fabrication non-idealities such as lithography induced non-uniformity across the sample.

\begin{figure*}[htbp]
\centering
\includegraphics[width=1\textwidth]{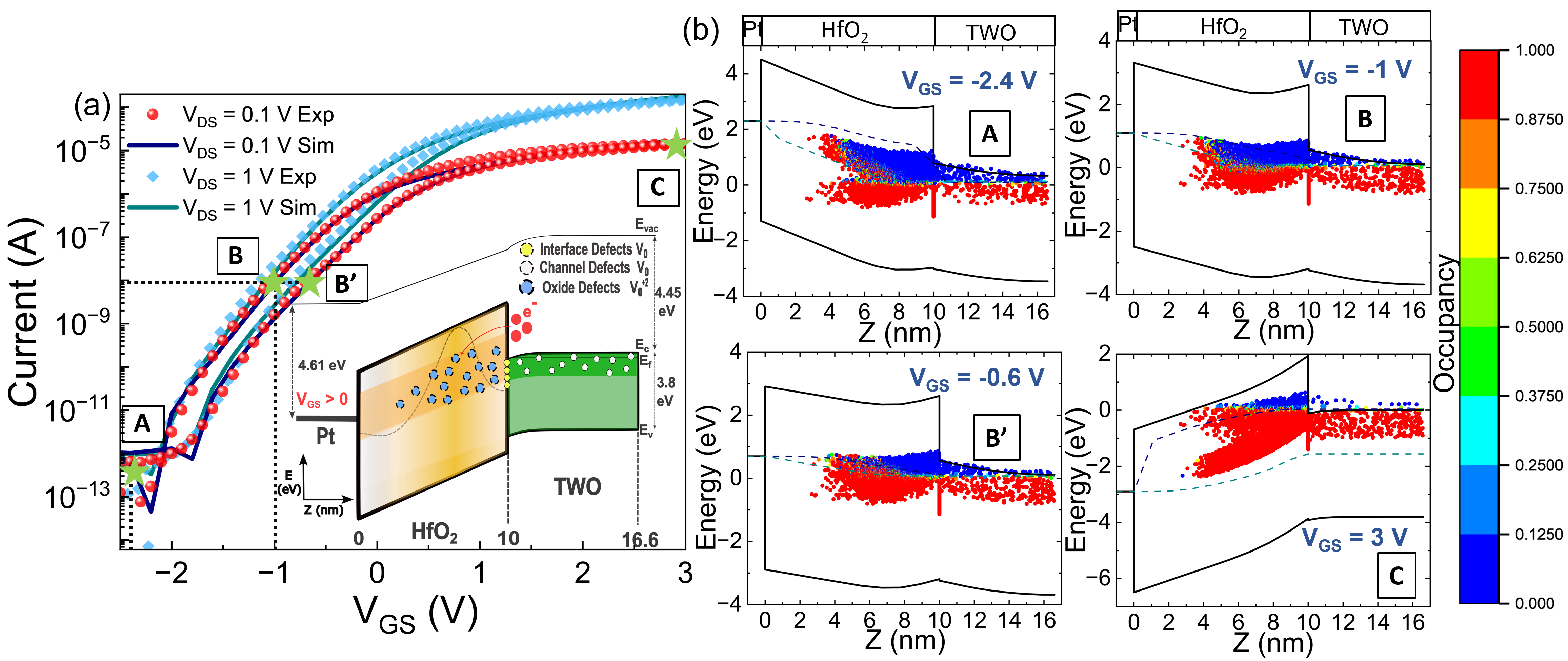}
\caption{\label{fig:Fig. 4}(a) Measured and Ginestra$^{\mathrm{TM}}$-simulated I$_{\mathrm{DS}}$ vs. V$_{\mathrm{GS}}$ characteristics at different drain voltages; the inset shows the assumed spatial and energy distribution of oxide defect states used in the simulation. (b) Simulated defect-state occupancy illustrating the origin of the hysteresis observed in (a). When the gate voltage is swept to 3 V (point C), deeper oxide traps become charged, leading to a shift in the V$_{\mathrm{th}}$.}
\end{figure*}
Charge trapping in the channel, at the gate dielectric/channel interface, and in the gate dielectric can significantly influence bias stress V$_{\mathrm{th}}$ instability as well as gate voltage hysteresis. To investigate this in detail kinetic Monte Carlo (kMC) simulations were carried out using Ginestra$^{\mathrm{TM}}$ TCAD software.~\cite{ref55} As illustrated in Fig.~\ref{fig:Fig. 4}(a), the simulated transfer characteristics closely follow the experimental forward as well as reverse gate voltage sweeps, showing good fit for V$_{\mathrm{DS}}$ of 0.1 V as well as 1 V. The defect distributions: a Gaussian profile of +2 charged oxygen vacancies V$_{\mathrm{o}}$$^{\mathrm{+2}}$ within the HfO$_{\mathrm{2}}$ layer\cite{ref37}, and a uniform distribution of neutral oxygen vacancies V$_{\mathrm{o}}$$^{\mathrm{0}}$ assumed in the upper part of the channel bandgap\cite{ref52,ref53,ref54} and at the gate dielectric/channel interface are depicted in the inset of Fig.~\ref{fig:Fig. 4}(a). The charge trapping mechanism is shown in Fig.~\ref{fig:Fig. 4}(b). Initially at point A (V$_{\mathrm{GS}}$ = $-2.4$ V), some oxide defect states are filled, while others remain empty. As the gate voltage increases to +3 V (point C), stronger band bending occurs at the gate dielectric/channel interface. This band bending allows accumulated channel electrons to tunnel into defect states located deeper inside the oxide, up to 4-6 nm from the gate dielectric/channel interface. When the gate voltage is reverse swept back to $-0.6$ V (point B$'$), not all trapped electrons are released immediately from the dielectric back into the channel. Some remain captured due to slow de-trapping kinetics. These residual trapped charges cause a positive shift in the V$_{\mathrm{th}}$ in the reverse sweep. The complete set of simulation parameters is listed in supplementary table S3. The channel doping concentration used in the Ginestra simulations was 1.5 × 10$^{19}$ cm$^{-3}$, which is comparable to the experimental value of 0.4 × 10$^{19}$ cm$^{-3}$ extracted from the analysis of transfer length method (TLM) measurements (Section S4 in supplementary material), further validating the model.

\begin{figure*}[htbp]
\includegraphics[width=0.8\textwidth]{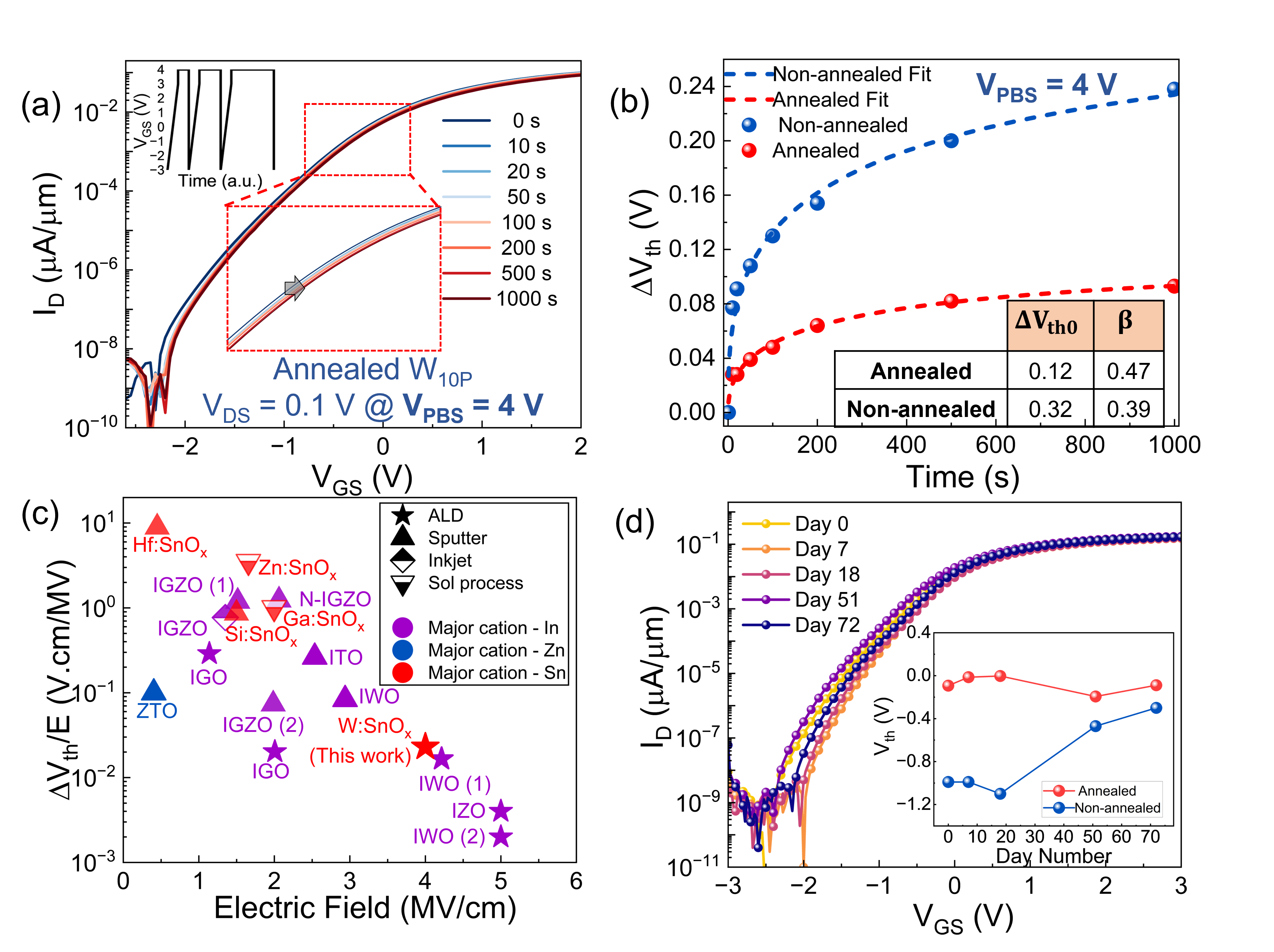}
\caption{\label{fig:Fig. 5}(a) Time-dependent I$_{\mathrm{DS}}$ vs. V$_{\mathrm{GS}}$ characteristics under positive V$_{\mathrm{GS}}$ vs. time stress (V$_{\mathrm{GS}}$ = 4 V) with applied gate bias pulse scheme shown in inset (top left) and magnified transfer curves shown in inset (center). (b) Threshold voltage shift ($\Delta$V$_{\mathrm{th}}$) as a function of PBS time comparing annealed and non-annealed devices, where solid symbols represent experimental data and the dashed lines represent fits using a stretched exponential model. Inset shows extracted $\Delta$V$_{\mathrm{th0}}$ and $\beta$ values for both devices. (c) Benchmarking of $\Delta$V$_{\mathrm{th}}$ normalized to the stress electric field ($\Delta$V$_{\mathrm{th}}$/E) for reported AOS TFTs highlighting major cation and deposition method of the channel oxide. (d) TFT transfer curve stability over time (days), with inset comparing annealed vs. non-annealed device V$_{\mathrm{th}}$ stability.}
\end{figure*}
Gate bias stress measurements were also performed to evaluate V$_{\mathrm{th}}$ stability of the TFTs. Fig.~\ref{fig:Fig. 5}(a) shows transfer characteristics of the annealed W$_{\mathrm{10P}}$ TFT measured under PBS with V$_{\mathrm{PBS}}$ = V$_{\mathrm{GS}}$ = 4 V applied to the gate terminal for stress times up to 1000 s. As the stress time increases, the transfer curves exhibit a small positive V$_{\mathrm{th}}$ shift, while the SS and the I$_{\mathrm{on}}$/I$_{\mathrm{off}}$ remain nearly unchanged. The positive V$_{\mathrm{th}}$ shift observed during PBS has been widely reported in oxide TFTs and is primarily associated with charge trapping phenomena.\cite{ref35} Under a 4 V gate bias, electrons tunnel into defect states and are captured by V$_{\mathrm{o}}$$^{\mathrm{+2}}$ located 4–6 nm inside the HfO$_{\mathrm{2}}$ layer (Fig.~\ref{fig:Fig. 4}(b)). The gradual increase in V$_{\mathrm{th}}$ with stress is attributed to electron trapping and their slow release from these distributed defect states, leading to a buildup of residual charge that increasingly screens the gate electric field and causes a progressive positive shift in V$_{\mathrm{th}}$. The threshold voltage shift ($\Delta$V$_{\mathrm{th}}$) is summarized in Fig.~\ref{fig:Fig. 5}(b) for both annealed and non-annealed devices. The time evolution of $\Delta$V$_{\mathrm{th}}$ follows a stretched exponential behavior (Section S7 in  supplementary material). The annealed device shows a much smaller shift of 0.093~V compared to 0.238~V for the non-annealed device. As shown in the inset of Fig.~\ref{fig:Fig. 5}(b), the extracted fitting parameters provide further insight into this improvement. The saturation shift ($\Delta$V$_{\mathrm{th0}}$), which represents the maximum threshold voltage shift, decreases from 0.32 to 0.12 V, indicating a reduction in trap density. The characteristic time constant ($\tau$) decreases from 473 s to 293 s, suggesting faster detrapping. Meanwhile, the stretching exponent ($\beta$), which reflects the spread of trap states, increases from 0.39 to 0.47, indicating a reduction in the dispersion of trap states, suggesting a more uniform and less disordered trapping environment. The reduced $\Delta$V$_{\mathrm{th}}$ observed after O$_{\mathrm{2}}$ annealing, together with the extracted fitting parameters, indicate effective passivation of oxygen vacancy defect states, leading to suppressed charge trapping and improved bias-stress stability.\cite{ref48,ref49,ref50} To benchmark the PBS stability, the normalized threshold voltage shift ($\Delta$V$_{\mathrm{th}}$/E) after 1000 s of PBS is compared as a function of electric field with previously reported oxide TFTs fabricated using different channel oxides and deposition methods (Fig.~\ref{fig:Fig. 5}(c)). The ALD TWO TFT exhibits superior PBS stability, outperforming all previously reported SnO$_{\mathrm{2}}$ and doped-SnO$_{\mathrm{2}}$ TFTs in terms of normalized threshold voltage shift.\cite{ref19,ref34,ref35,ref36} Similar to In-based channels, the ALD process gives better PBS performance for Sn-based oxide channels as compared to sputter deposition or solution-based processing.\cite{ref37,ref38,ref39,ref40,ref41,ref42,ref43,ref44,ref45,ref46,ref47} Further improvement in PBS stability can be achieved through passivation, fine-tuning of the W doping concentration, and annealing condition optimization. The transfer characteristics of the annealed W$_{\mathrm{10P}}$ TFT measured over 72 days show minimal degradation in drain current and subthreshold behavior (Fig.~\ref{fig:Fig. 5}(d)), indicating good device stability. The inset reveals that annealed devices exhibit significantly smaller threshold voltage shifts compared to non-annealed devices (supplementary material Fig. S6), highlighting the role of annealing in improving electrical stability. This is likely due to a reduction in surface and bulk defects leading to reduced oxygen and moisture adsorption and minimizing V$_{\mathrm{th}}$ variation. In contrast, the instability observed in non-annealed devices may arise from ambient adsorption at vacancy sites. Oxygen and moisture can form charged species at these sites, creating a surface depletion layer that can modulate the carrier concentration in the channel, consistent with previous reports on SnO$_{\mathrm{x}}$-based TFTs.\cite{ref51}
\begin{table}[htbp]
\centering
\caption{Benchmarking of ALD W:SnO$_2$ TFTs demonstrated in this work with reported sub-20 nm SnO$_2$ and doped-SnO$_2$ BEOL-compatible TFTs.}
\label{tab:1}
\renewcommand{\arraystretch}{1.3}
\setlength{\tabcolsep}{2pt}
\scriptsize

\begin{tabular}{c c c c c c c c c c}
\toprule
\rowcolor{apricot}
\textbf{Channel} & \textbf{Deposition} & \textbf{Thickness} & \textbf{Gate} & \textbf{L$_{\mathrm{ch}}$} & \textbf{SS} & \textbf{I$_{\mathrm{on}}$} & \textbf{I$_{\mathrm{off}}$} & \textbf{I$_{\mathrm{on}}$/I$_{\mathrm{off}}$} & \textbf{PBS} \\
\rowcolor{apricot}
\textbf{Material} & \textbf{Method} & \textbf{(nm)} & \textbf{Dielectric} & \textbf{($\mu$m)} & \textbf{(V/dec)} & \textbf{(A/$\mu$m)} & \textbf{(A/$\mu$m)} & \textbf{} & \textbf{$\Delta$V$_{\mathrm{th}}$/E (time)} \\
\midrule

SnO$_2$\cite{ref15}  & PVD & 4.5 & 40 nm HfO$_2$ & 50 & 0.11 & $2\times10^{-7}$ & $2\times10^{-14}$ & $10^{7}$ & -- \\

SnO$_2$\cite{ref16} & ALD & 16 & 200 nm SiO$_2$ & 200 & 5 & $10^{-8}$ & $10^{-13}$ & $10^{6}$ & -- \\

Si:SnO$_2$\cite{ref20} & Sputtering & 5 & 200 nm AlO$_x$:Nd & 20 & 0.21 & $10^{-6}$ & $<10^{-15}$ & $6\times10^{9}$ & 
0.57 V.cm/MV (5.4$\times10^{4}$ s) \\

W:SnO$_2$\cite{ref25} & Sputtering & 12 & 100 nm SiO$_2$ & 100 & 0.4 & $2\times10^{-7}$ & $2\times10^{-13}$ & $10^{6}$ & -- \\

\textbf{W:SnO$_2$ (This work)} & \textbf{ALD} & \textbf{6.8} & \textbf{10 nm HfO$_2$} &
\textbf{5} & \textbf{0.22} &
$>10^{-7}$ &
$<10^{-16}$ &
$>10^{9}$ &
\textbf{0.023 V.cm/MV ($10^{4}$ s)} \\

\bottomrule
\end{tabular}
\end{table}
The annealed W$_{\mathrm{10P}}$ TWO TFT performance is benchmarked with reported sub-20 nm SnO$_{\mathrm{2}}$ devices in Table ~\ref{tab:1}. The 6.8 nm thick ALD channel with a 10 nm HfO$_{\mathrm{2}}$ gate dielectric in this study achieves competitive I$_{\mathrm{on}}$/I$_{\mathrm{off}}$ > 10$^{9}$  and a steeper SS (0.22 V/dec) at low temperature (300 $^\circ$C) and scaled channel length of 5 $\mu$m compared to previous reports that relied on PVD or sputtering of thicker channel oxides and higher thermal budgets. Compared to reported SnO$_{\mathrm{x}}$ and SnO$_{\mathrm{x}}$-based TFTs, this work reports W:SnO$_{\mathrm{2}}$ TFTs with superior electrostatics along with BEOL-compatibility.
\section*{Conclusion}
\vspace{-15pt}
In conclusion, this work demonstrates low-temperature ALD-grown tungsten-doped SnO$_{\mathrm{x}}$ as a channel material for thin-film transistors and shows that it can serve as a practical indium-free alternative for oxide electronics. A comparative study of undoped SnO$_{\mathrm{x}}$, WO$_{\mathrm{x}}$, and different W-doped compositions, indicates that moderate tungsten incorporation (10\%) provides the best balance between conductivity and device stability. Controlled W doping helps suppress excessive carrier concentration while maintaining good switching behavior. Further, post-fabrication O$_{\mathrm{2}}$ annealing plays a key role in enhancing device performance, reliability and electrical stability. The noticeable reductions in SS, hysteresis, and PBS $\Delta$V$_{\mathrm{th}}$, along with a 100x increase in the I$_{\mathrm{on}}$/I$_{\mathrm{off}}$ ratio, suggest that controlling oxygen vacancies is critical for stable operation in these films. Ginestra TCAD simulations further validate the experimental findings. These results highlight the potential of TWO as a scalable and BEOL-compatible channel material. Further optimization of W doping concentration, interface engineering, annealing conditions and passivation strategies will be important for improving long-term reliability and enabling practical implementation.

See the supplementary material for detailed studies on targeted and experimentally achieved tungsten concentrations in Tungsten-doped SnO$_{\mathrm{x}}$ (TWO) films, XRD and AFM characterization of annealed W$_{\mathrm{10P}}$ films, gate leakage characteristics (I$_{\mathrm{GS}}$-V$_{\mathrm{GS}}$) of the TWO TFTs with varying tungsten doping concentration, gated-transfer length method (G-TLM) analysis of W$_{\mathrm{10P}}$ TFTs, Ginestra$^{\mathrm{TM}}$ TCAD simulation parameters for the channel and the dielectric layer, positive bias stress (PBS) stability of non-annealed W$_{\mathrm{10P}}$ TFT, stretched exponential model fitting of PBS induced $\Delta$V$_{\mathrm{th}}$, and long-term stability of non-annealed W$_{\mathrm{10P}}$ TFT.

The authors acknowledge the Indian Institute of Technology Bombay Nanofabrication Facility (IITBNF) for the usage of its device fabrication and characterization facilities and the TaRANG-sponsored research project funded by the Ministry of Electronics and Information Technology for funding this work. The authors thank Valerio Lunardelli, and Luca Larcher, from Applied Materials. This work was supported by the Ginestra Academic Program.

\section*{AUTHOR DECLARATIONS}
\vspace{-15pt}
The authors have no conflicts to disclose.

\section*{DATA AVAILABILITY}
\vspace{-15pt}
The data that support the findings of this study are available within the article and its supplementary material.

\vspace{-20pt}
\providecommand{\noopsort}[1]{}\providecommand{\singleletter}[1]{#1}%

\end{document}


\setcounter{table}{0}
\setcounter{figure}{0}
\begin{center}
    {\Large \textbf{Supplementary Information}}\\[6pt]
    
    {\large ALD W-Doped SnO$_{\mathrm{2}}$ TFTs for Indium-Free BEOL Compatible Electronics}\\[10pt]
    
    Mansi Anil Patil$^{1}$, Devarshi Dhoble$^{1}$, Shivaram Kubakaddi$^{1}$, Mamta Raturi$^{1}$, Marco A Villena$^{2}$, Gaurav Thareja$^{3}$, and Saurabh Lodha$^{1*}$\\[6pt]
    
    {\small
    $^{1}$Department of Electrical Engineering, Indian Institute of Technology Bombay, Mumbai 400076, India\\
    $^{2}$Department of Electronics and Computer Technology, Faculty of Sciences, University of Granada, Granada 18071, Spain\\
    $^{3}$Applied Materials Inc., Santa Clara, California, USA
    }\\[6pt]
    
    \texttt{slodha@ee.iitb.ac.in}
\end{center}
\vspace{10pt}

\section{Targeted and experimentally achieved tungsten concentrations in tungsten-doped SnO$_{\mathrm{x}}$ (TWO) films}
\begin{table}[H]
\centering
\caption{Targeted and achieved (extracted from XPS data) W\% for various compositions of TWO films.}
\begin{tabular}{cccccc}
\toprule
Sample & Target W\% & Ratio (Sn:W) & Material & Achieved W\% \\
\midrule
W$_{\mathrm{0P}}$   & 0   & 1:0  & SnO$_{\mathrm{x}}$       & --   \\
W$_{\mathrm{100P}}$ & 100 & 0:1  & WO$_{\mathrm{x}}$        & --   \\
W$_{\mathrm{5P}}$   & 5   & 11:1 & Sn$_{\mathrm{0.95}}$W$_{\mathrm{0.05}}$O$_{\mathrm{x}}$& 4.7  \\
W$_{\mathrm{10P}}$  & 10  & 5:1  & Sn$_{\mathrm{0.9}}$W$_{\mathrm{0.1}}$O$_{\mathrm{x}}$  & 9.8  \\
\bottomrule
\end{tabular}
\end{table}
Table S1 compares the targeted and experimentally achieved W concentrations in the deposited TWO films. The supercycle ratios for SnO$_{\mathrm{x}}$ and WO$_{\mathrm{x}}$ were chosen based on their respective growth-per-cycle (GPC) values (1.16 \AA/cycle and 0.66 \AA/cycle). The W\% values extracted from XPS are found to be close to the targeted stoichiometry. This confirms that the ALD supercycle approach provides precise control over dopant incorporation, which is critical for tuning electrical properties.

\section{XRD and AFM characterization of annealed W$_{\mathrm{10P}}$ films}
\begin{figure}[H]
\centering
\includegraphics[width=0.5\textwidth]{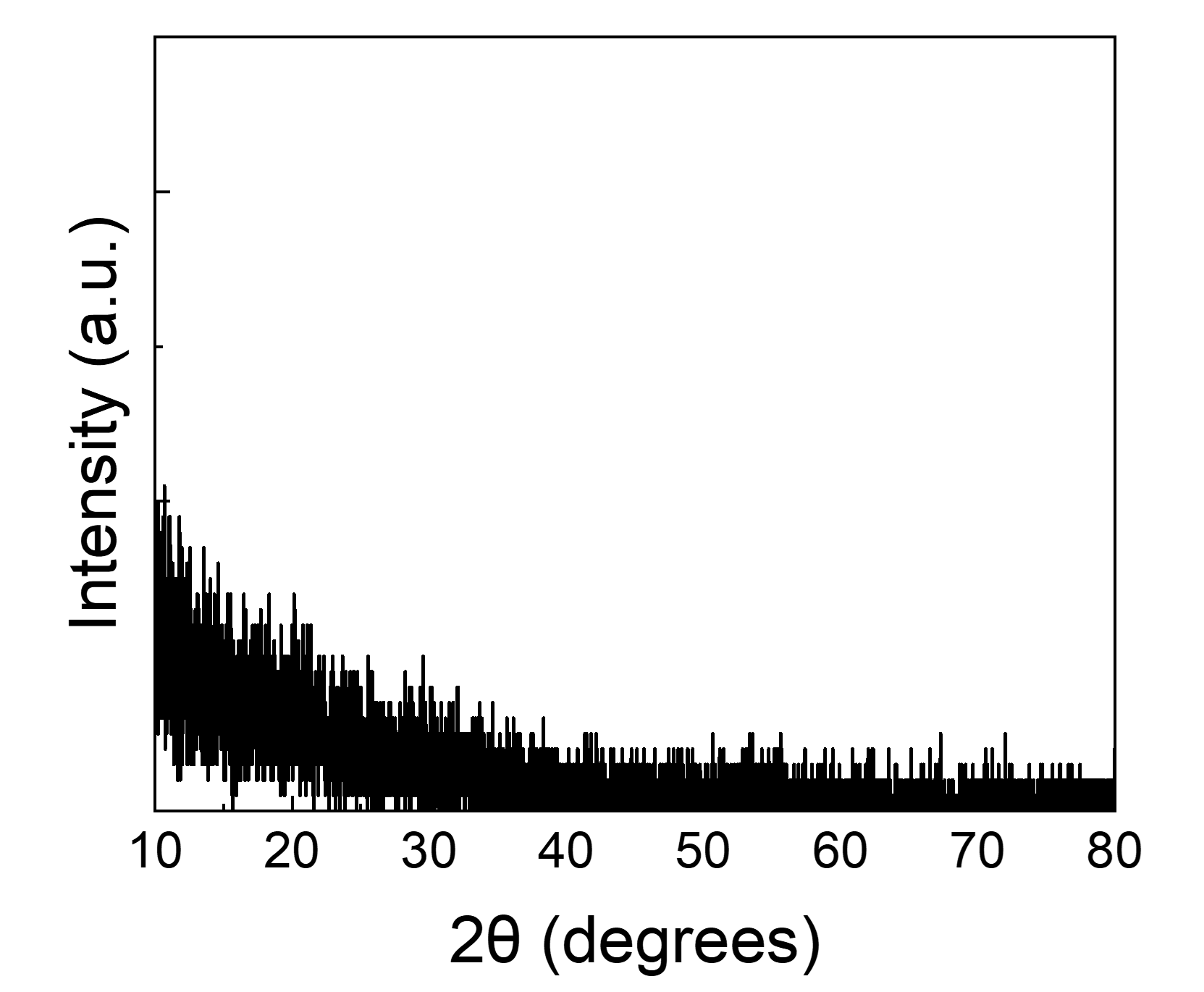}
\caption{X-ray diffraction (XRD) pattern of the annealed W$_{\mathrm{10P}}$ TWO film deposited on HfO$_{\mathrm{2}}$ substrate, showing the absence of distinct diffraction peaks and confirming the amorphous nature of the film.}
\end{figure}
\begin{figure}[H]
\centering
\includegraphics[width=\textwidth]{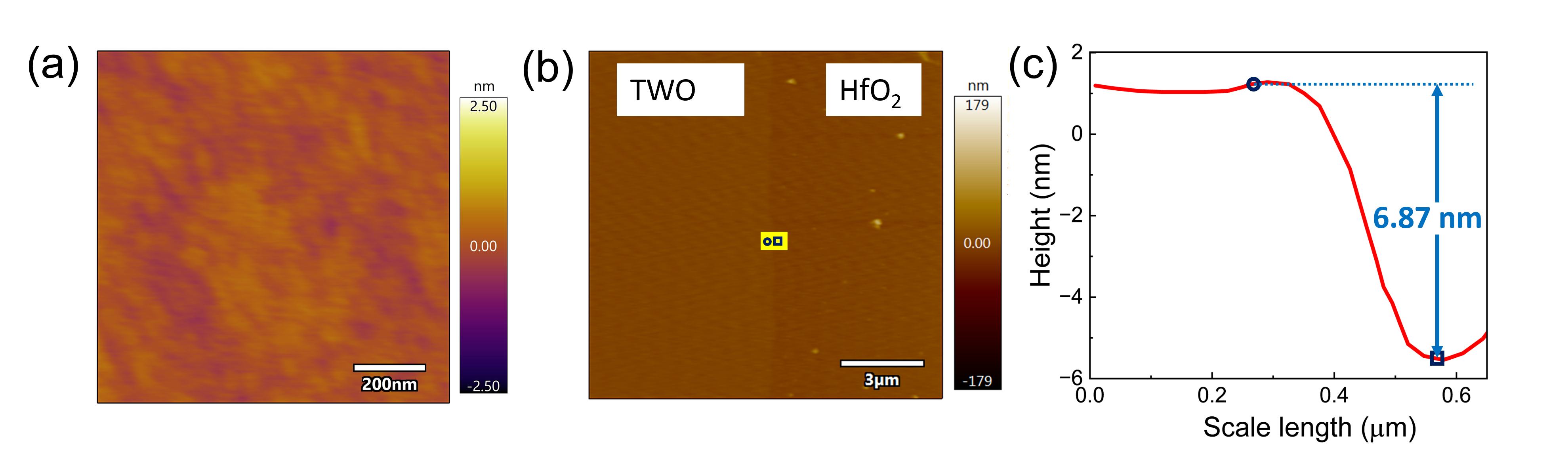}
\caption{Atomic force microscopy (AFM) characterization of the W$_{\mathrm{10P}}$ TWO film. (a) Surface topography scan over an area of 1 $\mu$m $\times$ 1 $\mu$m. (b) AFM topography of the annealed W$_{\mathrm{10P}}$ TWO film on HfO$_{\mathrm{2}}$; the yellow line indicates the position of the step height profile. (c) Step height profile extracted along the line in (b), indicating a TWO film thickness of $\sim$6.87 nm.}
\end{figure}
The XRD pattern of the annealed W$_{\mathrm{10P}}$ film (Fig. S1) does not show any distinct diffraction peaks, consistent with its amorphous nature. The broad background further supports the absence of long-range crystallinity in the deposited TWO layer.
Atomic force microscopy (AFM) measurements show a smooth film surface with an rms roughness of ~0.15 nm (Fig. S2 (a)). The step height of the TWO film deposited on HfO$_{\mathrm{2}}$ was extracted from a line scan across the film edge (yellow line in Fig. S2(b)), yielding a thickness of $\sim$6.87 nm (Fig. S2(c)), consistent with TEM and ellipsometry measurements discussed in the main text.

\section{Gate leakage characteristics (I$_{\mathrm{GS}}$-V$_{\mathrm{GS}}$) of the TWO TFTs with varying tungsten doping concentration}
\begin{figure}[H]
\centering
\includegraphics[width=0.6\textwidth]{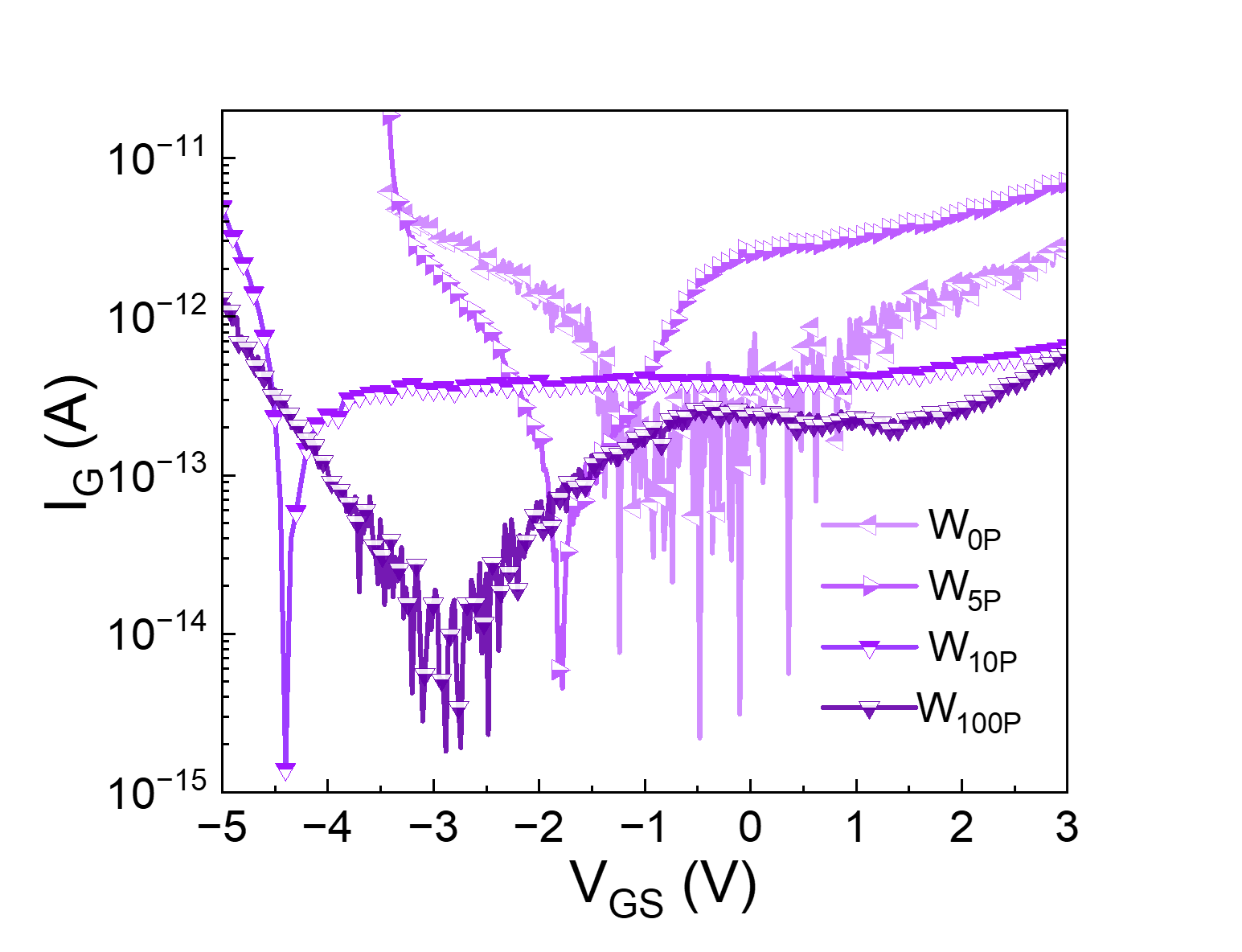}
\caption{Gate leakage characteristics (I$_{\mathrm{GS}}$ vs. V$_{\mathrm{GS}}$) of the TWO TFTs for varying tungsten doping concentrations.}
\end{figure}
The gate leakage current (I$_{\mathrm{GS}}$) as a function of gate voltage (V$_{\mathrm{GS}}$) for different W concentrations is shown in Fig. S3. The W$_{\mathrm{0P}}$ and W$_{\mathrm{5P}}$ devices start showing higher leakage at negative gate bias ($< -3$ V) and also exhibit relatively high I$_{\mathrm{off}}$. The high leakage can be linked to the high conductivity of W$_{\mathrm{0P}}$ and W$_{\mathrm{5P}}$ films, which makes it difficult to fully turn off the W$_{\mathrm{0P}}$ and W$_{\mathrm{5P}}$ TFTs with 6.8 nm channel thickness.\cite{ref1} In contrast, devices with higher W content (W$_{\mathrm{10P}}$ and W$_{\mathrm{100P}}$) show lower leakage, indicating better control over the channel.
\section{Gated-transfer length method (G-TLM) analysis of W$_{\mathrm{10P}}$ TFTs} 
\begin{figure}[H]
\centering
\includegraphics[width=\textwidth]{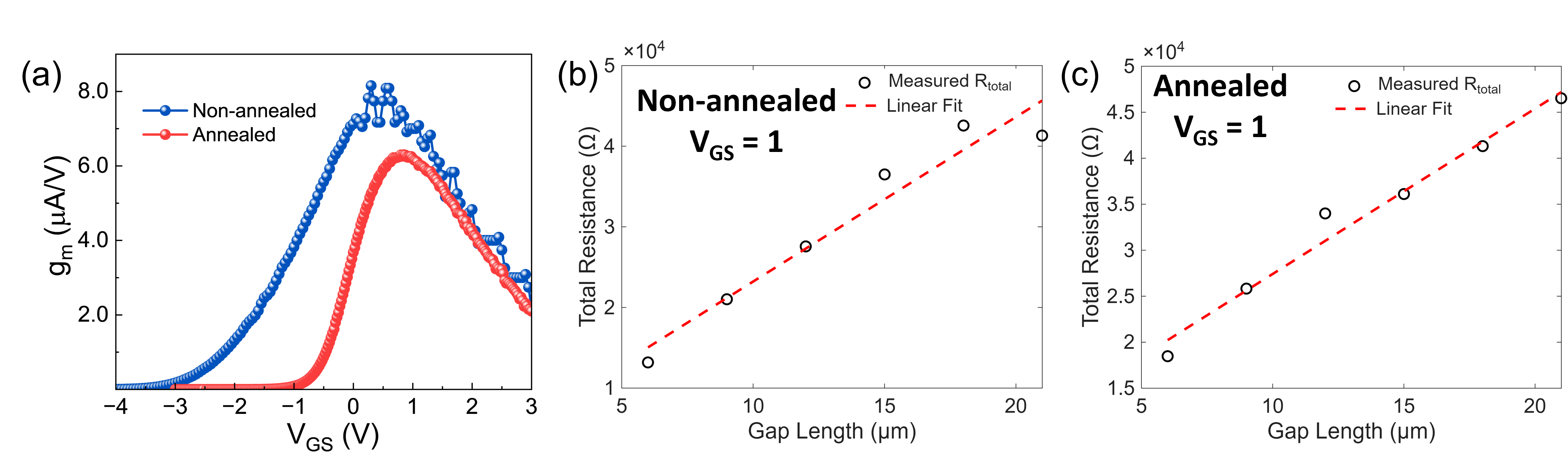}
\caption{(a) Transconductance (g$_{\mathrm{m}}$) as a function of gate voltage (V$_{\mathrm{GS}}$) for annealed and non-annealed devices. Total resistance as a function of contact spacing at V$_{\mathrm{GS}}$ = 1 V for (b) non-annealed and (c) annealed G-TLM structures.}
\end{figure}
\begin{table}[h]
\centering
\begin{tabular}{lcc}
\toprule
 & \textbf{Non-annealed} & \textbf{Annealed} \\
\midrule
R$_c$ ($\Omega$) & $1.4 \times 10^{3}$ & $4.7 \times 10^{3}$ \\
R$_s$ ($\Omega$/sq) & $102 \times 10^{3}$ & $89.8 \times 10^{3}$ \\
$\mu_{\mathrm{FE}}$ (cm$^{2}$/V$\cdot$s) & 5.8 & 6.6 \\
n$_s$ (cm$^{-2}$) & $9.1 \times 10^{12}$ & $2.9 \times 10^{12}$ \\
N$_{\mathrm{ch}}$ (cm$^{-3}$) & $1.3 \times 10^{19}$ & $0.4 \times 10^{19}$ \\
\bottomrule
\end{tabular}
\caption{Electrical parameters extracted from the G-TLM measurements for the non-annealed and annealed devices.}
\end{table}
G-TLM measurements were performed at V$_{\mathrm{GS}} = 1$ V to evaluate the contact and sheet resistance of the devices. After oxygen annealing, the sheet resistance decreases from 102 k$\Omega$/sq to 89.8 k$\Omega$/sq, while the contact resistance increases from 1.4 k$\Omega$ to 4.7 k$\Omega$. The reduction in sheet resistance, despite a decrease in carrier concentration caused by fewer oxygen vacancies, results from improved channel transport. Oxygen annealing lowers the density of subgap states, as reflected in the improvement of subthreshold swing (SS) from 410 to 220~mV/dec, corresponding to a decrease in interface trap density (D$_{\mathrm{it}}$) from $\sim 6.1 \times 10^{13}$ to $\sim 2.8 \times 10^{13}$~cm$^{-2}$eV$^{-1}$. Here, D$_{\mathrm{it}}$ was calculated from SS using
\[
\mathrm{SS} = \frac{\mathrm{k}\,\mathrm{T}}{\mathrm{q}} \ln 10 \left( 1 + \frac{\mathrm{C}_{\mathrm{it}}}{\mathrm{C}_{\mathrm{ox}}} \right),
\]
with $\mathrm{C}_{\mathrm{it}} = \mathrm{q}^2 \mathrm{D}_{\mathrm{it}}$. The reduced trap density suppresses trap-assisted scattering, improving carrier transport and enhancing the effective field-effect mobility. The increase in mobility exceeds the reduction in the carrier density, resulting in a net drop in the sheet resistance. The increase in contact resistance is due to lower carrier density near the metal–semiconductor interface, which limits current injection. The observed reduction in transconductance (g$_{\mathrm{m}}$) (Fig. S4(a)) after annealing therefore arises mainly from these contact limitations rather than from intrinsic degradation of channel transport. 
The field-effect mobility\cite{ref8}, sheet carrier density\cite{ref8}, and channel carrier concentration were determined using the following steps:
\begin{equation*}
\begin{aligned}
&\text{(1)}\quad
\mathrm{\mu}_{\mathrm{FE}} = \frac{1}{\mathrm{R}_{\mathrm{s}} \, \mathrm{C}_{\mathrm{ox}} \, (\mathrm{V}_{\mathrm{GS}} - \mathrm{V}_{\mathrm{th}})}\\
&\text{(2)}\quad
\mathrm{n}_{\mathrm{s}}  = \frac{\mathrm{L} \, \mathrm{I}_{\mathrm{DS}}}{\mathrm{q} \, \mathrm{W} \, \mathrm{\mu}_{\mathrm{FE}} \, \mathrm{V}_{\mathrm{DS}}} \\
&\text{(3)}\quad
\mathrm{N}_{\mathrm{ch}} = \frac{\mathrm{n}_{\mathrm{s}}}{\mathrm{t}_{\mathrm{ch}}}
\end{aligned}
\end{equation*}
Here, R$_{\mathrm{s}}$ is the sheet resistance, C$_{\mathrm{ox}}$ is the gate oxide capacitance per unit area, V$_{\mathrm{GS}}$ is the gate voltage, and V$_{\mathrm{th}}$ is the threshold voltage. I$_{\mathrm{DS}}$ is the drain current measured at V$_{\mathrm{GS}} = 1$ V and V$_{\mathrm{DS}} = 0.1$ V, while L and W represent the channel length and width of the TFT, respectively. q is the elementary charge ($1.6 \times 10^{-19}$ C), and t$_{\mathrm{ch}}$ is the TWO channel thickness.
\section{Ginestra$^{\mathrm{TM}}$ TCAD simulation parameters for the channel and the dielectric layer}
Material parameters used for device simulations in Ginestra$^{\mathrm{TM}}$ are listed for both the dielectric HfO$_{\mathrm{2}}$ and the TWO channel material in table S3. The bandgap of TWO (3.8 eV) was taken from reported experimental studies.\cite{ref2} Since detailed material parameters for TWO are limited in literature, several properties such as electron affinity (4.45 eV), dielectric permittivity (12), and effective mass (0.3m$_{\mathrm{o}}$) were selected based on reported values for SnO$_{\mathrm{2}}$, which has similar electronic characteristics.\cite{ref3,ref4} The specific heat capacity was estimated using reported thermodynamic data for SnO$_{\mathrm{2}}$.\cite{ref5} Overall, these parameters provide a reasonable approximation for modeling the electrical behavior of TWO-based devices. Table S4 presents the spatial and energy defect distribution parameters used for the TCAD simulations for both dielectric and channel materials.
\begin{table}[htbp]
\centering
\begin{tabular}{lcc}
\hline
\textbf{Properties} & \textbf{HfO$_{\mathrm{2}}$} & \textbf{TWO} \\
\hline
Band Gap & 5.8 eV & 3.8 eV \\
Electron Affinity & 2.4 eV & 4.45 eV \\
Relative Dielectric Permittivity & 21 & 12 \\
Specific Heat per Unit Volume & 1.164 J\,cm$^{-3}$K$^{-1}$ & 2.5 J\,cm$^{-3}$K$^{-1}$ \\
Thermal Conductivity & 0.005 W\,cm$^{-1}$K$^{-1}$ & 0.012 W\,cm$^{-1}$K$^{-1}$ \\
Electron DOS Effective Mass & 0.25$m_0$ & 0.3$m_0$ \\
Electron Tunneling Effective Mass & 0.25$m_0$ & 0.3$m_0$ \\
Hole DOS Effective Mass & 0.25$m_0$ & 1$m_0$ \\
Hole Tunneling Effective Mass & 0.25$m_0$ & 1$m_0$ \\
Electron Mobility & 1 cm$^2$V$^{-1}$s$^{-1}$ & 8 cm$^2$V$^{-1}$s$^{-1}$ \\
Electron Saturation Velocity & $2 \times 10^7$ cm\,s$^{-1}$ & $1 \times 10^7$ cm\,s$^{-1}$ \\ 
Hole Mobility & 1 cm$^2$V$^{-1}$s$^{-1}$ & 1 cm$^2$V$^{-1}$s$^{-1}$ \\
Hole Saturation Velocity & $2 \times 10^7$ cm\,s$^{-1}$ & $1 \times 10^7$ cm\,s$^{-1}$ \\
Phase & Amorphous & Amorphous \\
\hline
\end{tabular}
\caption{Material properties of HfO$_{\mathrm{2}}$ and TWO used in device modelling of a W$_{\mathrm{10P}}$ TFT in Ginestra$^{\mathrm{TM}}$ TCAD simulation.}
\end{table}
\begin{table*}[htbp]
\centering
\scriptsize
\setlength{\tabcolsep}{3pt}
\renewcommand{\arraystretch}{1.2}

\resizebox{\textwidth}{!}{
\begin{tabular}{c c cc cc c cc c cc cc}
\toprule

\multirow{3}{*}{$\phi_m$ (eV)} & 
\multicolumn{5}{c}{\textbf{HfO$_{\mathrm{2}}$ (Properties of dielectric defects)}} & 
\multicolumn{3}{c}{\textbf{HfO$_{\mathrm{2}}$/TWO interface defects}} & 
\multicolumn{5}{c}{\textbf{TWO (Properties of channel defects)}} \\

\cmidrule(lr){2-6} \cmidrule(lr){7-9} \cmidrule(lr){10-14}

& \multirow{2}{*}{$D_{ox}$ (cm$^{-3}$)} 
& \multicolumn{2}{c}{Physical Distribution} 
& \multicolumn{2}{c}{Energy Distribution} 
& \multirow{2}{*}{$D_{it}$ (cm$^{-2}$)} 
& \multicolumn{2}{c}{Energy Distribution} 
& \multirow{2}{*}{$N_D$ (cm$^{-3}$)} 
& \multicolumn{2}{c}{Physical Distribution} 
& \multicolumn{2}{c}{Energy Distribution} \\

& 
& $\mu$ (nm) & $\sigma$ (nm) 
& $\mu$ (eV) & Spread (eV) 
& 
& $\mu$ (eV) & Spread (eV) 
& 
& $\mu$ (nm) & Spread (nm) 
& $\mu$ (eV) & Spread (eV) \\

\midrule

4.61 & 
$5.5 \times 10^{19}$ & 
7.5 & 1.25 & 
2.2 & 1 & 

$3.32 \times 10^{13}$ & 
2.88 & 0.85 & 

$1.5 \times 10^{19}$ & 
13.3 & 6.6 & 
1 & 0.5 \\

\bottomrule
\end{tabular}
}
\caption{Spatial and energy distribution parameters of defects in HfO$_{\mathrm{2}}$, HfO$_{\mathrm{2}}$/TWO interface, and TWO channel.}
\end{table*}

\section{Positive bias stress (PBS) stability of non-annealed W$_{\mathrm{10P}}$ TFT}
\begin{figure}[H]
\centering
\includegraphics[width=0.6\textwidth]{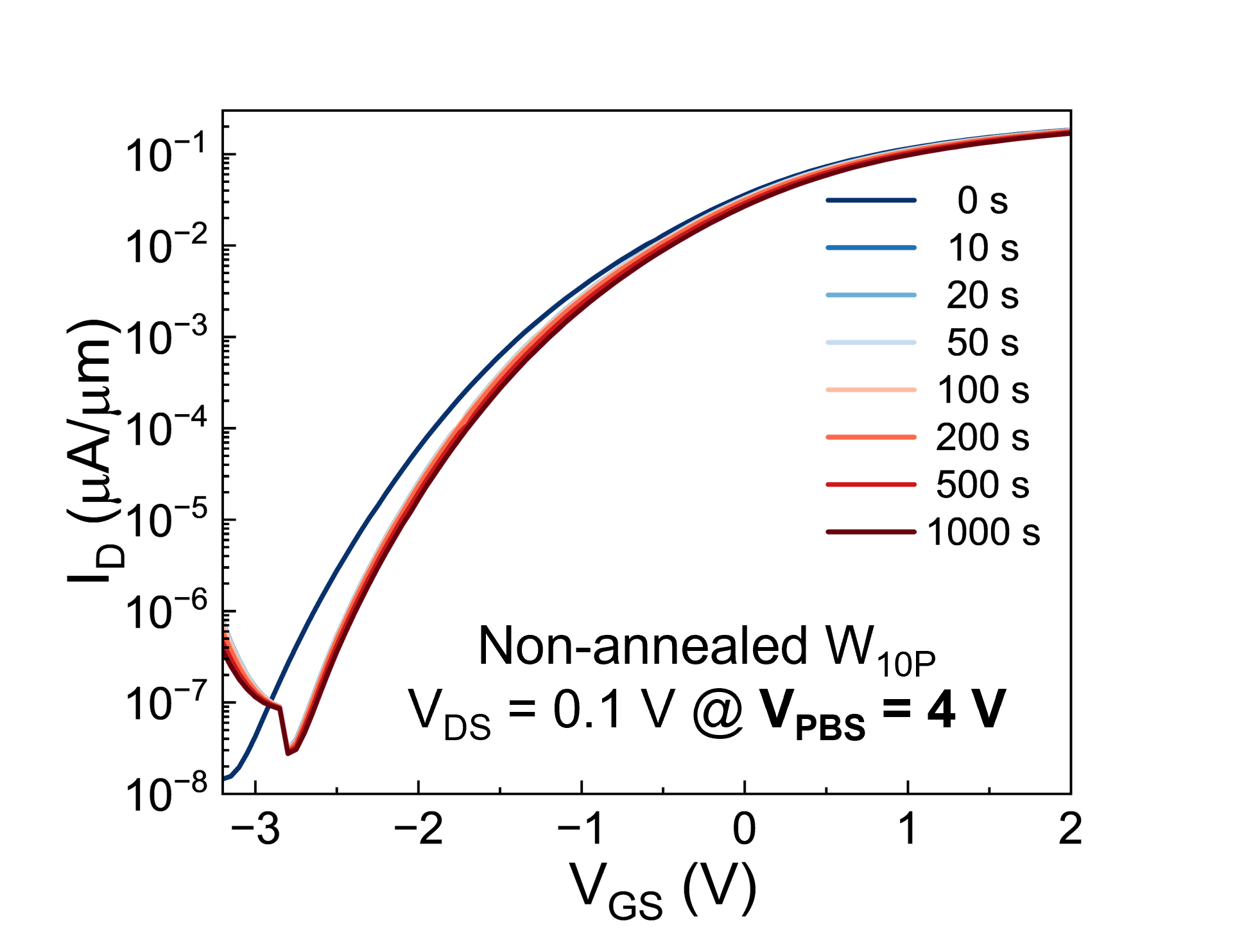}
\caption{Time-dependent transfer characteristics (I$_{\mathrm{DS}}$ vs. V$_{\mathrm{GS}}$) of non-annealed W$_{\mathrm{10P}}$ TWO TFT under positive bias stress (PBS) conditions (V$_{\mathrm{GS}}$ = 4 V).}  
\end{figure}
The non-annealed device shows a noticeable shift in the transfer characteristics (V$_{\mathrm{th}}$ shift of 0.238 V) under positive bias stress (V$_{\mathrm{GS}}$ = 4 V) for 1000s. Annealing in O$_{\mathrm{2}}$ ambient passivates defect states in the gate dielectric and at the channel/dielectric interface, reducing charge trapping compared to the non-annealed device. Therefore, the $\Delta$V$_{\mathrm{th}}$ shift is higher for the non-annealed TFT compared to the annealed TFT.
\section{Stretched exponential model fitting of PBS induced $\Delta$V$_{\mathrm{th}}$}
The time-dependent threshold voltage shift was analyzed using a stretched exponential model, commonly used to describe dispersive charge trapping dynamics in oxide semiconductors.\cite{ref6,ref7} Fitting model used:
\begin{equation*}
\mathrm{\Delta V}_{\mathrm{th}}(\mathrm{t}) = \mathrm{\Delta V}_{\mathrm{th0}}
\left[ 1 - \exp \left( -\left( \frac{\mathrm{t}}{{\tau}} \right)^{{\beta}} \right) \right]
\end{equation*}

\begin{table}[H]
\centering
\caption{Fitting parameters for stretched exponential model fitting of the $\Delta$V$_{\mathrm{th}}$ with respect to time for the annealed and non-annealed W$_{\mathrm{10P}}$ TFTs.}
\begin{tabular}{lcc}
\toprule
Parameter & Annealed & Non-annealed \\
\midrule
$\Delta$V$_{\mathrm{th}}$ (V) & 0.122 & 0.317 \\
$\tau$ (s) & 293 & 473 \\
$\beta$ & 0.47 & 0.39 \\
\bottomrule
\end{tabular}
\end{table}

\begin{itemize}

\item $\mathrm{\Delta V}_{\mathrm{th}}$: Final saturation value of the threshold voltage shift. 
After annealing, $\Delta$V$_{\mathrm{th0}}$ decreases from 0.317 V to 0.122 V.

\item $\tau$: Characteristic time constant representing the trapping timescale. 
It decreases from 473~s to 293~s after annealing, indicating faster de-trapping and reduced V$_{\mathrm{th}}$ shift.

\item $\beta$: Stretching exponent describing the distribution of trap states ($0 < \beta < 1$). 
A value of $\beta = 1$ corresponds to uniform trapping, whereas $\beta < 1$ indicates a broad distribution of trap energies.

\end{itemize}
The increase in $\beta$ from 0.39 to 0.47 after annealing indicates a reduction in the dispersion of trap states, suggesting a more uniform and less disordered trapping environment.
\section{Long-term stability of non-annealed W$_{\mathrm{10P}}$ TFT}
\begin{figure}[H]
\centering
\includegraphics[width=0.6\textwidth]{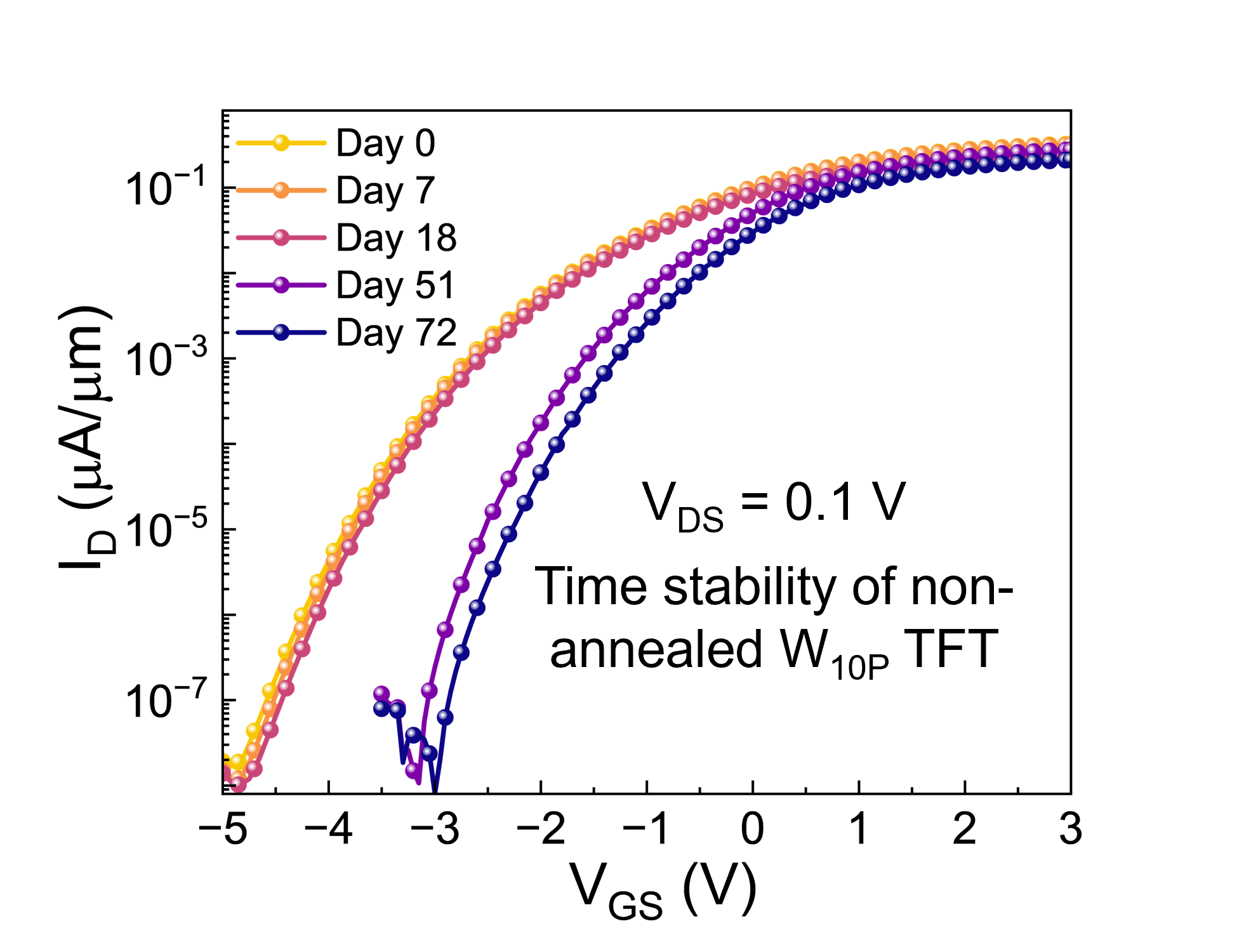}
\caption{Long-term stability of non-annealed W$_{\mathrm{10P}}$ TFT measured over 72 days.}
\end{figure}
The long-term stability of non-annealed devices was evaluated by monitoring device characteristics over several days.  An overall shift of around 0.7 V was observed over 72 days, indicating gradual changes in device properties during ambient exposure.

\providecommand{\noopsort}[1]{}\providecommand{\singleletter}[1]{#1}%